\documentclass[12pt]{iopart}
\usepackage{graphicx}
\usepackage{bm}
\usepackage{iopams}

\begin{document}

\def\star{{\displaystyle *}}
\def\be{\begin{equation}}
\def\ee{\end{equation}}

\title[Solid state plasmas]{Solid state plasmas}

\author{Giovanni Manfredi and J\'{e}r\^{o}me Hurst}
\address{Institut de
Physique et Chimie des Mat\'{e}riaux, CNRS and
Universit\'{e} de Strasbourg, BP 43, F-67034 Strasbourg, France}
\ead{Giovanni.Manfredi@ipcms.unistra.fr}


\begin{abstract}
Magnetic fusion devices operate at regimes characterized by extremely high temperatures and low densities, for which the charged particles motion is well described by classical mechanics. This is not true, however, for solid-state metallic objects: their density approaches $10^{28} \rm m^{-3}$, so that the average interparticle distance is shorter than the de Broglie wavelength, which characterizes the spread of the electron wave function. Under these conditions, the conduction electrons behave as a true quantum plasma even at room temperature.
Here, we shall illustrate the impact of quantum phenomena on the electron dynamics in metallic objects of nanometric size, particularly thin metallic films excited by short laser pulses. Further, we will discuss more recent results on regimes that involve spin and relativistic effects.
\end{abstract}

\maketitle

\section{Introduction} \label{sec:intro}
Attempts at reproducing on Earth, in a controlled way, the nuclear fusion reactions that fuel most stars, including the Sun, have mainly followed two approaches -- magnetic confinement fusion (MCF) and inertial confinement fusion (ICF). Both approaches rely on a mixture (plasma) of electrons and hydrogen ions (deuterium and tritium), which at very high temperatures have a finite probability to fuse together to yield heavier ions (helium), thereby releasing large amounts of energy in the form of kinetic energy of the reaction products (helium ions and neutrons).

At a fundamental level, these two competing approaches differ primarily in the
physical features of the plasmas. A simple measure for a fusion reactor to reach ignition is provided by the Lawson criterion, which gives a minimum required value for the ``triple product" of the plasma density $n$, the plasma temperature $T$, and the energy confinement time $\tau_E$:
\(
nT\tau_E > 3 \times 10^{21}~ \rm keV ~m^{-3}~ s.
\)
Broadly speaking, MCF plasmas correspond to low densities ($n\simeq 10^{20} \rm m^{-3}$) and long confinement times ($\tau_E \simeq 1\rm s$), whereas ICF plasmas display large densities ($10^{30}-10^{32} \rm m^{-3}$) and short confinement times ($\tau_E \simeq 10^{-10}\rm s$).  In both cases, the temperature needs to be around 100 million degrees ($T \simeq 10 \rm keV$) in order for the fusion reactions to occur.
These densities should be compared to that of atmospheric air, basically a perfect gas ($10^{25} \rm m^{-3}$) and solid metal objects ($10^{28} \rm m^{-3}$). Thus, MCF plasmas are much more rarefied than ordinary gases, while ICF plasmas are even denser than solids.

Can one expect any quantum mechanical phenomena to take place in such plasmas? Certainly not for MCF. We know that quantum features occur at high densities (short distances) and low temperature. Since atmospheric air behaves as a classical gas at room temperature, then {\em a fortiori} this should remain so for the more rarefied and hotter plasmas occurring in MCF. For ICF plasmas, the situation is more ambiguous, as they are both denser and hotter than ordinary solid matter and we know that electrons in solids behave quantum mechanically at room temperature. As we shall see, ICF plasmas are usually on the border between the classical and the quantum regimes.

While quantum effects have no impact on the {\it dynamics} of MCF plasmas, they do play an important role in determining the cross sections of the fusion reactions. For the reactions to take place, the hydrogen ions need to overcome the repulsive Coulomb barrier, which is about 300\,keV.
However, quantum tunneling allows the classical barrier to be overcome at somewhat lower energies, of the order of 100\,keV. Then, for a plasma with temperature $T \simeq 10\, \rm keV$, there are enough particles in the Maxwellian tail with sufficient energy to tunnel through the Coulomb barrier and trigger a reaction. We will not consider this aspect here and rather concentrate on the influence of quantum effects on the collective plasma motion.

This review will focus on the quantum electron dynamics in solid, particularly metals, for which the conduction electrons can be viewed as a mobile plasma neutralized by the background ions. Quantum effects arise because of the large density, which means that electrons are closely packed together. These quantum features become even more apparent for metallic objects of nanometric size ($\rm 1\,nm = 10^{-9}\, m$), such as metallic nanoparticles, thin films, and nanorods, which have stimulated a huge amount of scientific interest in the last two decades \cite{Bigot_annphys}, both for fundamental research (they are at the border between classical and quantum physics) and for potential technological applications that range from physical chemistry, to biology and medicine.
For instance, ``plasmonic resonances", ie, electronic oscillations near the plasma frequency $\omega_p=\sqrt{e^2n/m\varepsilon_0}$, are routinely observed in metallic nanoparticles and their properties (resonance width, damping, dipole and quadrupole modes, \dots) are studied experimentally using ultrafast spectroscopy techniques. Indeed, as the plasma period in metallic nanoparticles is of the order of one femtosecond ($10^{-15}$\,s), the recent development of femto- and attosecond laser sources has opened up a vast domain of research that is still being explored \cite{Stockman,Scholl}.

Finally, quantum plasma effects can be observed in astrophysical systems -- interior of giant planets, white dwarfs, neutron stars and pulsars -- due to the extreme conditions of density, temperature, and magnetic fields that exist in such environments \cite{Uzdensky}.

Here, we will first review some of the basic aspects of solid-state electron plasmas, with particular emphasis on kinetic descriptions. These methods will help us illustrate the impact of quantum phenomena on the electron dynamics in metallic nano-objects, particularly thin metallic films excited by ultrashort laser pulses.
Secondly, we will present new theoretical advances related to recent experiments \cite{Bigot} on ferromagnetic thin films, where the magnetic degrees of freedom of the electron (spin and orbital angular momentum)
can play an important part. Finally, we will briefly hint at relativistic effects, which also have an impact on the spin dynamics.

\section{Basic concepts} \label{sec:basic}
The basic tenets of quantum plasma physics have been presented in previous reviews \cite{manfredi2005, manfredi2009} and will be briefly summarized here.

A fermion gas at equilibrium obeys the Fermi-Dirac (FD) distribution. At zero temperature, all energy levels are occupied up to the Fermi energy
\be
E_F = \frac{\hbar^2}{2m}~(3\pi^2)^{2/3}~ n^{2/3}~,
\label{fermi-energy}
\ee
which is a function of the electron mass $m$ and density $n$ ($\hbar$ is the reduced Planck constant). One can also define the Fermi temperature $T_F=E_F/k_B$, where $k_B$ is the Boltzmann constant, and the Fermi velocity $v_F=\sqrt{2E_F/m}$. A convenient parameter to quantify the degree of ``quantumness" of an electron gas is the {\em degeneracy parameter} $\chi=T/T_F$: when $\chi \gg 1$ the FD distribution tends to a Maxwell-Boltzmann one and the gas behaves classically; in contrast, $\chi \ll 1$ describes the fully quantum regime.

The degeneracy parameter can also be expressed in terms of the thermal de Broglie wavelength $\lambda_B = \hbar/\sqrt{m k_B T}$, which is a measure of the spread of the electron wave function. One can easily show that $\chi \sim (d/\lambda_B)^2$, where $d=n^{-1/3}$ is the average interparticle distance. Quite naturally then, quantum effects become important when the electron wave functions overlap significantly.

Two further important dimensionless quantities are the coupling parameters $g$, which characterize the degree of collisionality of the plasma. They can be expressed as the ratio of the interaction (Coulomb) energy $E_{\rm coul}=e^2 /(\varepsilon_0 d)$ to the typical kinetic energy. In the classical regime, the latter is given by the thermal energy, so that the coupling parameter is the usual one:
\be
g_C = \frac{E_{\rm coul}}{k_B T}= \frac{e^2 n^{1/3}}{\varepsilon_0 k_B T} \sim \left( \frac{1}{n\lambda_D^3}\right)^{2/3}, \label{gc}
\ee
where we have introduced the Debye length $\lambda_D=\sqrt{k_B T \varepsilon_0/en}$.
In the deep quantum regime, the typical kinetic energy is the Fermi energy and the coupling parameter becomes:
\be
g_Q = \frac{E_{\rm coul}}{E_F} =  \frac{2}{(3\pi^2)^{2/3}}~
\frac{e^2 m }{\hbar^2 \varepsilon_0 ~n^{1/3}} \sim \left(
\frac{1}{n \lambda_F^3}\right)^{2/3}  ~, \label{gq}
\ee
where $\lambda_F=v_F/\omega_p$ is the Thomas-Fermi screening length, the quantum analogue of the Debye length. A classical or quantum plasma is collisionless (ie, weakly coupled) when the relevant coupling parameter is much less than unity.
In condensed matter physics, models that are valid when $g \ll 1$ are often referred to as {\em mean-field models}.

\begin{table}
\centering \caption{Typical time, space, velocity, and energy scales for
bulk gold and ICF plasmas.}
\medskip
\label{tab:1}       
\begin{tabular}{llll}
\hline\noalign{\smallskip}
 & Solid gold & ICF & Units  \\
\noalign{\smallskip}\hline\noalign{\smallskip}
$n$ & $5.9 \times 10^{28}$ & $10^{32}$ & $\rm m^{-3}$  \\
$T$ & 300 & $10^{8}$ &\rm K \\
$E_F$ & 5.53  & 785 & \rm eV \\
$T_F$ & 64200 & $9 \times 10^{6}$ &\rm K \\
$\lambda_F$ & 0.1 & 0.03 & \rm nm \\
$d=n^{-1/3}$ & 0.25 & 0.022 & \rm nm \\
$v_F$ & $1.4 \times 10^{6}$ & $1.7 \times 10^{7}$ & $\rm ms^{-1}$ \\
$2\pi\omega_{p}^{-1}$ & 0.46 & 0.01 & fs \\
$\hbar \omega_{p}$ & 9.02 & 371 & eV \\
$g_Q$ & 12.7 & 1.07 & ---\\
$\chi$ & $4.7 \times 10^{-3}$ & 11 & ---\\
\noalign{\smallskip}\hline
\end{tabular}
\end{table}

In order to fix the ideas, let us consider gold nanoparticles, which are typical metallic nano-objects routinely used in the experiments. The typical time, space, and energy scales for gold are summarized in Table \ref{tab:1}. Note that these values are meaningful at thermodynamic equilibrium and for bulk macroscopic matter.
First, we notice that the Fermi temperature is very high, therefore $\chi \ll 1$ even at room temperature: electrons in solid metals are always degenerate and behave quantum-mechanically. Second, the coupling parameter $g_Q$ is larger than unity, ie, electron correlations must play a role. Collisionless (mean field) approaches may be fine to understand qualitatively the electron dynamics, but in order to obtain quantitative results more sophisticated models are required.
We also see that the typical time, space, and energy scale are given respectively by the femtosecond, the nanometer, and the electron-volt.
Finally, as the Fermi velocity is much less than the speed of light $c=3\times 10^8 \rm\, m/s$, we do not expect any relativistic effects, at least at equilibrium, although strong electromagnetic pulses may accelerate the electrons out of equilibrium to relativistic velocities.

For comparison, Table \ref{tab:1} also shows the values for typical ICF experiments. The most important difference is that ICF plasmas are on the border of degeneracy ($\chi \simeq 10$); in other words, contrarily to MCF plasmas, they can display weak quantum effects in their dynamics, although, because of their high temperature, they are still very far from the fully degenerate regime ($\chi \ll 1$). We also see that ICF plasmas are closer to the collisionless regime, as their coupling parameter is close to unity (note that when $\chi \approx 1$, then $g_C \approx g_Q$).

\section{Plasmon resonances in nano-objects -- Mie theory} \label{sec:plasmon}
In most current experiments, nano-objects are excited via ultrashort laser pulses with a pulse duration that can go down to a few hundred attoseconds. Femtosecond or longer pulses have been almost routine for the last two decades. The wavelength of the radiation usually lies in the visible range (400--800\,nm), although x-ray and infrared pulses are also envisageable (although less easily produced). Thus, the laser wavelength is much longer than the size of the nano-objects and the electromagnetic fields can be viewed as spatially uniform inside the object (dipole approximation).

\begin{figure}[t]
    \begin{center}
    \includegraphics[width=.5\linewidth]{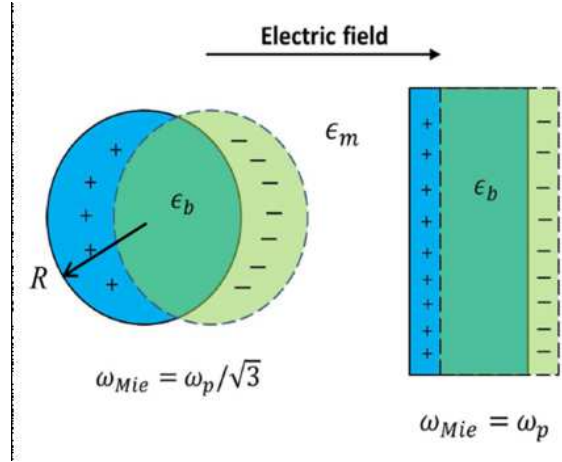}
    \end{center}
    \caption{\footnotesize Schematic representation of the plasmon resonance for a spherical nanoparticle (left panel) and a planar film (right panel).}
    \label{fig:plasmon}
\end{figure}

In this approximation, the laser electric field pulls the conduction electrons away from the more massive ions, thus initiating self-consistent oscillations of the electron gas. At resonance, when the frequency of the external electric field equals the natural frequency of the electron gas in the nano-object, the absorption cross-section reaches a maximum. Using purely classical arguments based on Maxwell's equations and considering spherical nanoparticles, the resonant frequency turns out to be the Mie frequency \cite{kreibig}:
\be
\omega_{\rm Mie} = \frac{\omega_p}{\sqrt{2\epsilon_m + \epsilon_b}},
\label{Mie-freq}
\ee
where $\epsilon_m$ is the dielectric constant of the environment where the nanoparticle is embedded and $\epsilon_b$ is the dielectric constant of the bound electrons inside the particle (see Fig. \ref{fig:plasmon}, left panel). Taking $\epsilon_b=\epsilon_m=1$ for simplicity, yields $\omega_{\rm Mie}=\omega_p/\sqrt{3}$. The factor $\sqrt{3}$ comes from the spherical symmetry that we assumed. For a planar film (Fig. \ref{fig:plasmon}, right panel), the resonant frequency is simply: $\omega_{\rm Mie}=\omega_p$.

\begin{figure}[t]
    \begin{center}
    \includegraphics[width=.5\linewidth]{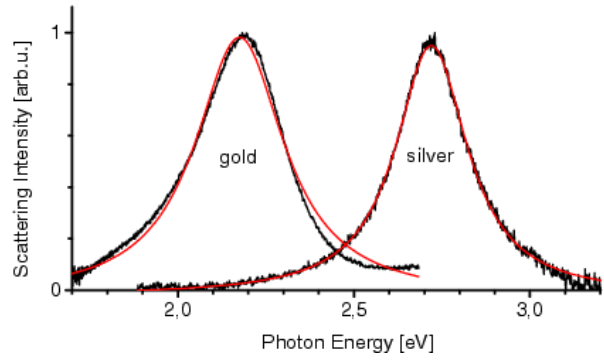}
    \end{center}
    \caption{\footnotesize Scattering spectra of gold and silver spherical nanoparticles with diameter 60\,nm (from Ref. \cite{sonnichsen}).}
    \label{fig:gold-silver}
\end{figure}

An example of measured scattering spectrum for gold and silver nanoparticles is shown in Fig. \ref{fig:gold-silver}. For spherical gold particles, the Mie frequency in energy units is $\hbar \omega_{\rm Mie}=5.2\,\rm eV$ (see Table \ref{tab:1}), which differs significantly from the value on Fig. \ref{fig:gold-silver}, although the order of magnitude is correct. The discrepancy come from various sources of damping, which tend to red shift the resonant frequency, as in a damped oscillator. The damping rate $\Gamma$ is given by the linewidth of the resonance curve: in this case $\Gamma \approx 0.4\rm \, eV$, which is rather large compared to the observed resonant frequency $\omega_{\rm res} \approx 2.2\rm \, eV$. Nevertheless, a proper application of the Mie theory, including the realistic dielectric constants and various forms of damping, reproduces the correct value of the resonance frequency and linewidth \cite{sonnichsen}.

In metallic nano-objects, the main sources of damping are electron-electron (e-e) and electron-phonon (e-ph) collisions, as well as radiation damping. When $T \ll T_F$, e-e collisions are strongly suppressed, because almost all energy levels below $E_F$ are full and there are no available states for the scattered electrons to occupy. This effect is known as {\em Pauli blocking} \cite{manfredi2005,Ashcroft}.
E-ph collisions can be crudely estimated using Drude's classical theory \cite{Ashcroft}. For gold at room temperature, the e-ph relaxation time is around $\Gamma_{\rm e-ph}^{-1}\approx 28 \rm \,fs$, which in energy units yields $\Gamma_{\rm e-ph}\approx 0.15 \rm \,eV$, and accounts for almost half of the damping observed in Fig. \ref{fig:gold-silver}.
Radiation damping occurs because the oscillating electrons behave as an electric dipole, thus emitting electromagnetic waves. The damping rate can be estimated by computing the total power radiated by the dipole \cite{Jackson}. One obtains \be
\frac{\Gamma_{\rm rad}}{\omega_{\rm res}} \approx \frac{e^2}{2\varepsilon_0hc}
~\frac{\hbar \omega_{\rm res}}{mc^2}~N \approx \frac{1}{137}~\frac{\hbar \omega_{\rm res}}{511\, \rm keV}~N,
\label{radiation}
\ee
where $N$ is the the number of electrons in the nanoparticle. For the 60-nm diameter particles of Fig. \ref{fig:gold-silver}, one obtains approximately $\Gamma_{\rm rad}\approx 0.46 \rm \,eV$, which is actually the dominant source of damping in this case.

\section{Kinetic models} \label{sec:models}
The use of kinetic phase-space models for the electron dynamics in metallic nano-objects was described in detail in previous reviews of ours \cite{manfredi2005,manfredi2009}.
An electron plasma constitutes a many-particle system that in principle should be described by the $N$-body Schr\"{o}dinger equation. However, for more than a few electrons, this task is computationally untractable, hence the need of approximate models. In condensed-matter theory, the ``mother-of-all-approximations" is given by the time-dependent Hartree equations (TDHEs). This is the analogue of the Vlasov-Poisson equations (VPEs) in plasma physics: it retains the self-consistent electric field (mean field), but neglects e-e correlations. Just like the VPEs, the TDHEs are a good approximation when the coupling parameter is small. This is well-established for most fusion plasmas ($g_C \ll 1$), but not so much for electrons in metals where $g_Q \geq 1$. For this reason, a battery of improvements on the Hartree equations has been developed, which can be grouped in two categories:
\begin{itemize}
\item
The Hartree-Fock equations retain a purely quantum kind of e-e correlations, called {\em exchange}, which results from the anti-symmetric character of the $N$-body wave function for fermions;
\item
Density functional theory (DFT) can in principle accommodate all e-e correlations in the form of extra potentials that depend on the electron density. In practice, such potentials are only known approximately. Thus, DFT can be view as a formal ``exactification" \footnote{This is the expression used by Walter Kohn, the founder of DFT, in his 1998 Nobel lecture \cite{kohn-nobel}.} of the Hartree theory.
\end{itemize}

The TDHEs read as a set of one-body Schr\"{o}dinger equations coupled to Poisson's equation for the electric potential $V({\bm r},t)$:
\begin{eqnarray}
i\hbar\frac{\partial\psi_\alpha}{\partial\,t} &=& -
\frac{\hbar^2}{2m}~ \Delta \psi_\alpha
- eV\psi_\alpha~, ~~~~~ \alpha = 1 \dots N \label{eq:hartree}\\
\Delta V &=& \frac{e}{\varepsilon}\left(\sum_{\alpha=1}^N p_\alpha |\psi_{\alpha}|^2 - n_{i}(\bm{r})\right),
\label{eq:poisson}
\end{eqnarray}
where $n_{i}(\bm{r})$ is the ion density and $p_\alpha$ is the occupation probability for the state $\psi_\alpha$.
The link between the quantum TDHEs and the classical VPEs can be made through the Wigner transformation \cite{Wigner,Hillery}
\be
f(\bm{r},\bm{v},t) = \sum_{\alpha=1}^N \frac{m}{2\pi\hbar}~
p_\alpha \int_{-\infty}^{+\infty} \psi_\alpha^{\displaystyle*}
\left(\bm{r} + \frac{\bm s}{2},t\right) \psi_\alpha\left(\bm{r} -
\frac{\bm s}{2},t\right) e^{i m {\bm v} \cdot {\bm s}/\hbar}~d{\bm s}.
\label{wigfunc}
\ee
Using the Wigner transformation, the TDHE can be written in the form of a phase-space evolution equation (Wigner equation)
\[
\frac{\partial{f}}{\partial{t}} + {\bm v}\cdot \nabla f+ \frac{em}{2i\pi\hbar^2}
\]
\be
\times \int\int{d{\bm s}}~{d{\bm v}'}e^{im({\bm v}-{\bm v}')\cdot{\bm s}/\hbar}
\left[V\left({\bm r}+\frac{\bm s}{2}\right)-
V\left({\bm r}-\frac{\bm s}{2}\right)\right]f({\bm r},{\bm v}',t) = 0, \label{wignereq}
\ee
where the electric potential obeys Poisson's equation (\ref{eq:poisson}).

It can be shown that Wigner's evolution equation (\ref{wignereq}) formally reduces to the Vlasov equation when $\hbar \to 0$. The Wigner formalism thus constitute a useful tool to compare directly the classical and quantum dynamics for the same physical system. We will do just that in the next section for the case of a thin metallic film.

\section{Electron dynamics in thin metal films} \label{sec:metalfilms}
As an illustration, we present here some of the results that were obtained by our group along a period of several years \cite{manf-herv,jasiak,jasiak-deco}.
We model our metal film as a slab of thickness $L$ in the $x$ direction and much larger extension in the transverse plane (see Fig. \ref{fig:plasmon}, right panel). The ions constitute an immobile background of uniform positive charge, with density $n_0$ inside the slab and zero outside. In this configuration, the motion of an electron in the transverse plane is decoupled from the motion normal to the surface of the film and a one-dimensional (1D) model along $x$ can be adopted.

\begin{figure}
\centering
\includegraphics[width=0.4\textwidth]{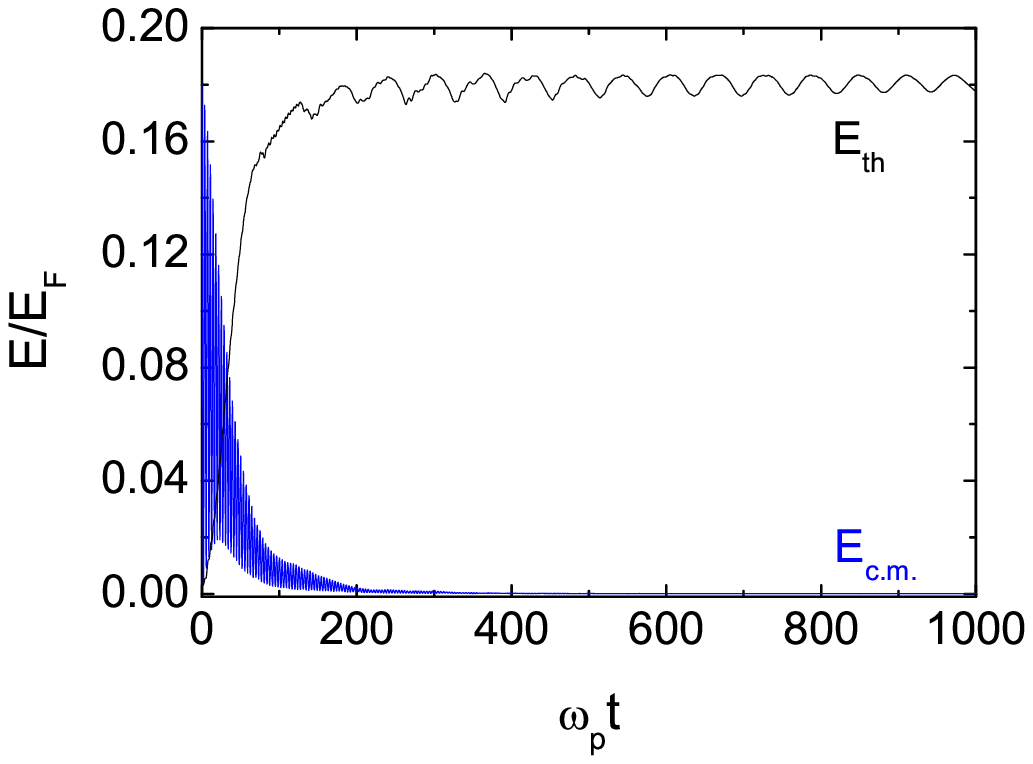}
\includegraphics[width=0.4\textwidth]{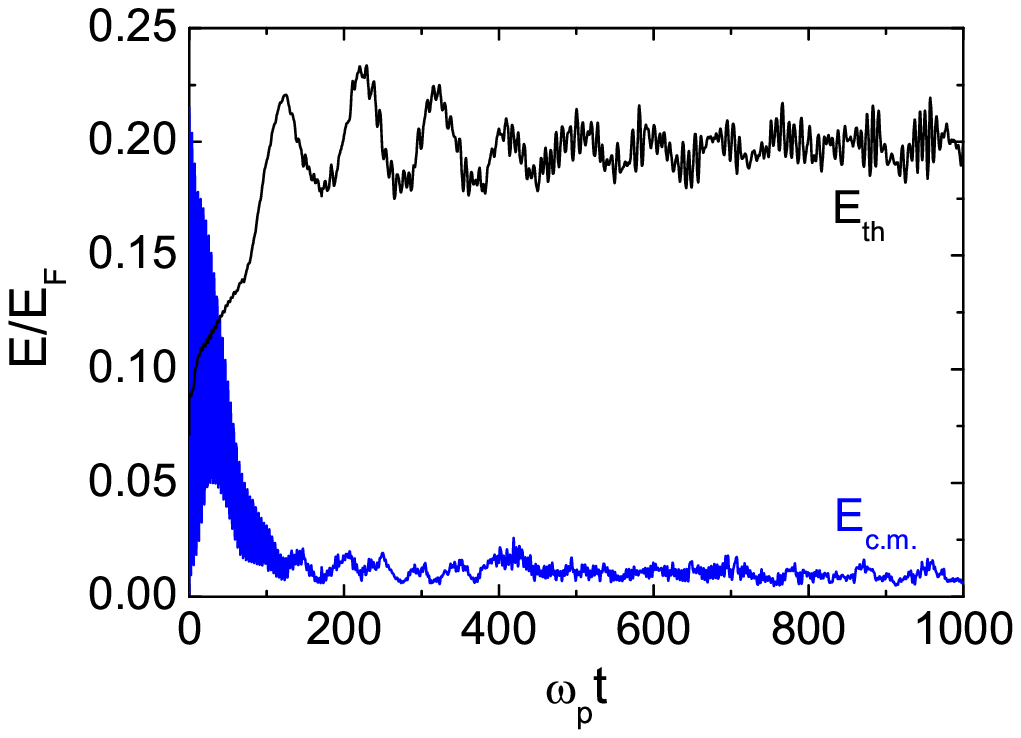}
\caption{Evolution of the thermal and centre-of-mass energies
in the Vlasov (left panel) and Wigner (right
panel) cases, for a sodium film of thickness $L=6\,\rm nm$ (from Ref. \cite{jasiak-deco}).}
\label{fig:ekin006}
\end{figure}

The electrons are initially prepared in a
FD equilibrium at finite temperature. They
are subsequently excited by imposing a constant velocity shift
$\delta v$ to the initial distribution. The electron dynamics is computed by solving numerically the Vlasov or Wigner equations on a phase-space grid. In particular, we have analyzed the time evolution of the thermal energy $E_{\rm th}$ and the center-of-mass kinetic energy $E_{\rm cm}$ (Fig. \ref{fig:ekin006}). During an initial rapidly-oscillating phase,
$E_{\rm cm}$ is almost entirely converted into thermal energy through
Landau damping. After saturation, a slowly oscillating regime
appears. The period $T$ of these oscillations is very close to the time of flight between the film surfaces for electrons traveling at the Fermi velocity. Thus, such oscillations correspond to bunches of electrons bouncing back and
forth on the film surfaces, as one could verify by directly inspecting the phase space portraits \cite{manf-herv}.

\begin{figure}[!ht]
\centering
\includegraphics[width=0.4\textwidth]{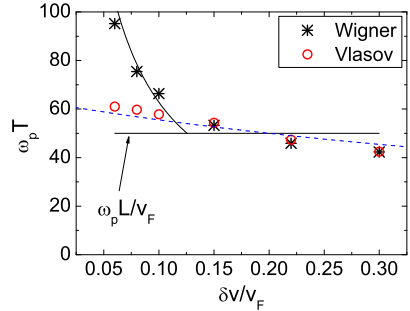}
\caption{Period of the low-frequency oscillations of Fig.~\ref{fig:ekin006} as a
function of the perturbation. The horizontal line represents the classical time of flight (from Ref. \cite{jasiak-deco}).}
\label{fig:persl50T300}
\end{figure}

However, the period of these oscillations is not quite the same when we use the Vlasov or the Wigner approach, as can be seen from Fig.~\ref{fig:ekin006}. By repeating the simulations for different excitations, it turns out that the classical and quantum results coincide for strong excitations, but diverge for small ones (Fig. \ref{fig:persl50T300}). Thus, we observe a clear transition between a classical and a
quantum regime occurring at a fairly well-defined threshold. The threshold corresponds to an excitation with energy equal to the plasmon energy $\hbar\omega_p$. These results provide a practical example of a quantum-classical transition in the electron dynamics that could, in principle, be observed experimentally.

\begin{figure}
\centering
\includegraphics[width=0.6\textwidth]{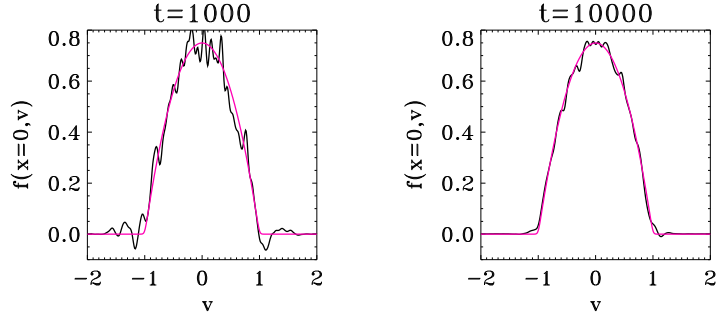}
\caption{Velocity distribution at the midpoint of the film, for two different times measured in units of $\omega_p$. The red line represents the 1D equilibrium FD distribution.}
\label{fig:fv}
\end{figure}

An important advantage of the Wigner formalism (compared to the Schr\"{o}dinger formalism of DFT) is that collisional effects can be more easily be added to the model, by analogy with the classical Fokker-Planck (FP) equation. In order to model e-ph collisions in our thin film dynamics, we have added a FP term
on the right-hand side of the Wigner equation (\ref{wignereq}): $(\partial_t f)_{\rm e-ph} = D \nabla^2_v f + \gamma \nabla_v \cdot({\bf v}~ G[f])$,
where $\gamma$ is the nominal relaxation rate, $D$ is a diffusion coefficient in velocity space, and $G[\cdot]$ is a functional that depends on the quantum statistics and on the dimensionality of the system \cite{jasiak-deco}.
A judicious choice of $G[\cdot]$ yields that
$(\partial{f}/\partial{t})_{\rm e-ph} =0$ when the electrons follow a FD distribution in 1D.
This approach enabled us to study the approach to equilibrium after an external excitation, which had been neglected in the preceding analysis. The results are shown in Fig. \ref{fig:fv}, where we plot a cut of the Wigner function $f$ against the velocity $v$, at the midpoint of the film. Under the action of the FP term, $f$ tends to its FD equilibrium. During the evolution, the Wigner function becomes everywhere positive, so that it can be again interpreted as a true probability density in the phase space. This process, whereby quantum correlations are lost to an external environment (the phonon bath), constitutes the essence of {\em decoherence}.

\section{Spin and relativistic effects} \label{sec:spinrelat}
\subsection{Spin Vlasov equations}
The electron carries not only an electric charge, but also a spin. In recent years, there has been a surge of interest in the spin dynamics in solid-state devices \cite{Bigot_annphys}, as a possible means to store and transport information (spintronics), as well as for attempts at developing quantum computing devices \cite{Zoller}.

In order to take into account the electron spin at a nonrelativistic level, the Schr\"{o}dinger wave function must be replaced by a 2-spinor
\be
\Psi_\alpha (\bm{r},t) =
\left(\begin{array}{c}
\psi_\alpha^\uparrow (\bm{r},t)  \\
\psi_\alpha^\downarrow (\bm{r},t)  \end{array} \right). \label{spinor}
\ee
Then, the TDHEs are replaced by the mean-field Pauli equations:
\be
i\hbar \frac{\partial \Psi_{\alpha}}{\partial t} =
\left[ \frac{1}{2m}(-i\hbar\bm{\nabla}+e\bm{A})^2 -eV \right] \Psi_{\alpha}+
\mu _{B} \bm{\sigma} \cdot \bm{B}\, \Psi_{\alpha},
\label{pauli_eq}
\ee
where $\mu_B=e\hbar/2m$ is Bohr's magneton, $\bm{\sigma}$ are the Pauli matrices, and $\bm{B}=\bm{\nabla}\times \bm{A}$. The term $\mu _{B} \bm{\sigma} \cdot \bm{B}$ represents the Zeeman effect. The scalar and vector potentials $V$ and $\bm{A}$ can be either external or self-consistent.

The Wigner function that corresponds to the spinor (\ref{spinor}) is a $2\times 2$ matrix:
\be
\mathcal{F}(\bm{r},\bm{v},t) =
\left(\begin{array}{cc}
f^{\uparrow \uparrow} & f^{\uparrow \downarrow}  \\
f^{\downarrow \uparrow} & f^{\downarrow \downarrow}\end{array} \right). \label{f-matrix}
\ee
It is convenient to project $\mathcal{F}$ onto the Pauli basis \cite{barletti_03}, such that:
$
\mathcal{F}= \frac{1}{2}f_0 + \frac{1}{\hbar}\bm{f} \cdot \bm{\sigma},
$
where
\begin{equation}
f _{0} = \textrm{tr} (\mathcal{F})  = f ^{\uparrow \uparrow} + f ^{\downarrow \downarrow}, ~~~~
\bm{f}   = \frac{\hbar}{2} \textrm{tr} \left( \mathcal{F} \bm \sigma \right)
\label{paulibasis}
\end{equation}
and tr denotes the trace. Now, $f_0$ is the analogue of the ordinary phase-space distribution, while $\bm{f}$ (with components $f_{i}, ~i=x,y,z)$ is related to the spin polarization in the direction $i$. In other words, $f_{0}$ represents the probability to find an electron at one point of the phase space at a given time, whereas $f_{i}$ represents the probability to have a spin polarization probability in the direction $ \hat {\bm e}_i$ for that electron.

The evolution equations obeyed by the Wigner functions (\ref{paulibasis}) are rather complicated. In the limit $\hbar \to 0$, they reduce to the following set of Vlasov equations \cite{vlasovia}
\begin{eqnarray}
\frac{\partial f_{0}}{\partial t} &+& \bm{v} \cdot \bm{\nabla}f_{0}
- \frac{e}{ m } \left(\bm{E} + \bm{v} \times \bm{B} \right) \cdot \bm{\nabla}_{\bm{v}} f_{0} - \frac{e }{m^{2}} \sum_{i} \bm{\nabla} B_{i} \cdot \bm{\nabla}_{\bm{v}} f_{i} = 0,  \label{f0_evo_vlasov}  \\
\frac{\partial f_{i}}{\partial t} &+& \bm{v} \cdot \bm{\nabla}f_{i}
- \frac{e}{ m } \left[ \left(\bm{E} + \bm{v} \times \bm{B} \right)\cdot \bm{\nabla}_{\bm{v}} f_{i} - \left( \bm{f} \times \bm{B} \right) _{i} \right] - \frac{\mu_B \hbar}{2m}  \bm{\nabla} B_{i} \cdot \bm{\nabla}_{\bm{v}} f_{0}.
 =0, \label{falpha_evo_vlasov}
\end{eqnarray}
Both the electron charge and spin are subject to the Lorentz force; in addition the spins precess around the magnetic field ($\bm{f} \times \bm{B}$ term). Charges and spins are coupled via a term that depends on the gradient of the magnetic field (last term in both equations). Equations (\ref{f0_evo_vlasov})-(\ref{falpha_evo_vlasov}) constitute an intermediate model where the electron motion is classical, while the spin is treated as a fully quantum variable.

Some authors have used an alternative representation based on a single scalar Wigner function (instead of a $2\times 2$ matrix) evolving in an extend phase space: $F\left(\bm{r},\bm{v},\widehat{\bm{s}} \right)$, where $\widehat{\bm{s}}$ is a vector of unit length \cite{zamanian_NJP10}. The two approach are mathematically equivalent, as one can go from the extended phase-space distribution $F$ to the matrix Wigner function $\mathcal{F}$ through a simple linear transformation \cite{vlasovia}.

\begin{figure}
\centering
\includegraphics[width=0.5\textwidth]{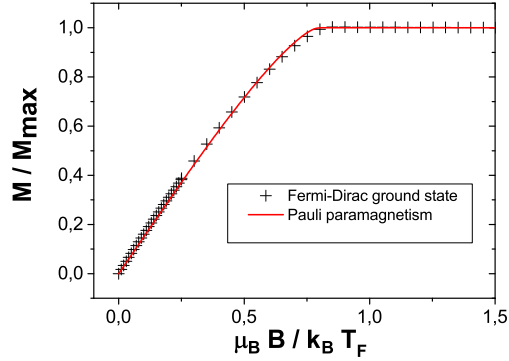}
\caption{Relative electron magnetization as a function of the external magnetic field normalized to the Fermi energy, for a Fermi-Dirac equilibrium at temperature $T=300~$K. Symbols represent the numerical results, while the red solid line is the theoretical curve for Pauli paramagnetism.}
\label{fig:pauli}
\end{figure}

\subsection{Stationary states}
We consider a 1D slab geometry as in the preceding section, with only variations in the $x$ direction taken into account. The external magnetic field $\bm{B}=B_z\bm{e}_z$ is uniform and parallel to $z$. We also suppose that the electrons can only be polarized along $z$ (collinear magnetism).
Therefore, at equilibrium one has: $f_{0} =  f_{0}(x,|\bm{v}|)$, $f_{z} =  f_{z}(x,|\bm{v}|)$, and $f_{x} = f_{y} =0$.

For the standard (spinless) Vlasov equation, the stationary states are functions of the Hamiltonian $H=mv^2/2-eV$. In our case, it is natural to take a FD equilibrium:
$F_{D} (H) = n_{0} [ 1 + \exp ((H-\mu)/k_B T)]^{-1}$, where $\mu$ is the chemical potential. When the spin is included, the Vlasov equations (\ref{f0_evo_vlasov})-(\ref{falpha_evo_vlasov}) can be written as (braces denote Poisson's brackets):
\be
  \frac{\partial f_{0}}{\partial t} =
 \left\{  H^{\uparrow \uparrow} , f^{\uparrow \uparrow}  \right\} +  \left\{  H^{\downarrow \downarrow} , f^{\downarrow \downarrow}  \right\},~~~ \frac{\partial f_{z}}{\partial t} =  \left\{  H^{\uparrow \uparrow} , f^{\uparrow \uparrow}  \right\} - \left\{  H^{\downarrow \downarrow} , f^{\downarrow \downarrow}  \right\},
\ee
where
$ H ^{\uparrow \uparrow} = \frac{m}{2}\bm{v}^{2}  + V + \mu_{B} B_{z}$ and $H^{\downarrow \downarrow} =\frac{m}{2}\bm{v}^{2}  + V - \mu_{B} B_{z}$.
We deduce that $f^{\uparrow \uparrow}$ at equilibrium must be a function of $H^{\uparrow \uparrow}$ and $f^{\downarrow \downarrow}$ a function of $H^{\downarrow \downarrow}$. Using the FD distribution, the stationary solutions are given by:
$
f_{0}^{stat} = F_{D} ( H^{\uparrow \uparrow} ) +  F_{D} ( H^{\downarrow \downarrow} )$ and $f_{z}^{stat} =  F_{D} ( H^{\uparrow \uparrow} ) - F_{D} ( H^{\downarrow \downarrow} )$.
Finally, the stationary state is found by computing the electron density $n=\int f_0 d\bm{v}$ and solving the resulting nonlinear Poisson equation to obtain the potential $V$. This is enough to specify the self-consistent FD equilibria for $f_0$ and $f_z$.

A uniform magnetic field has no impact on the electron dynamics [as only the gradients of $\bm{B}$ enter Eqs. (\ref{f0_evo_vlasov})-(\ref{falpha_evo_vlasov})], but affects the equilibrium, because it acts differently on spin-up and spin-down electrons. In Fig. \ref{fig:pauli}, we show numerical results for the total magnetization $M=\int\int f_z dx d\bm{v}$ as a function of the external magnetic field. Clearly, the magnetization is significantly different from zero only when $\mu_B B_z \ge k_B T_F$. For solid gold, this means $B_z \ge 10^5\,\rm T$, which is a huge magnetic field. This is consistent with the fact that Pauli's spin paramagnetism is very small at equilibrium \cite{Ashcroft,Acharyya}, since, for small temperatures, it is proportional to $(T/T_F)^2$.

The above result sheds some light on a recent controversy concerning spin fluid models, which are obtained by taking velocity moments of kinetic equations such as (\ref{f0_evo_vlasov})-(\ref{falpha_evo_vlasov}). For instance, the spin polarization is defined as $\bm{S}(\bm{r},t)=\int \bm f (\bm{r},\bm{v}, t) d\bm{v}$ and should be small when $\bm f$ is a FD equilibrium, as was correctly recognized in \cite{chippy,bonitz}. The problem with the fluid models is that the FD distribution is somewhat forgotten in the moment-taking procedure, so that it appears (incorrectly) that the spin polarization $\bm{S}$ may take any values at equilibrium. The authors of \cite{chippy,bonitz} conclude that the problem lies with the Hartree approximation, because it neglects the antisymmetric character of the $N$-body wave function. But this is too strong a statement. As we have seen, one can still use the Hartree approximation, provided that the equilibrium is a FD distribution. This is sufficient to yield the correct (and small) value of the spin polarization.

We also point out that the Pauli spin polarization is small because it involves only electrons with an energy close to the Fermi energy. However, these are the very electrons that are important for all dynamical phenomena (see, for instance, the phase space portraits in Ref.~\cite{manf-herv}), whereas electrons situated well below the Fermi level play virtually no role. Therefore, even though only a fraction of the electron density is polarized, it may still have a significant impact on the transport properties.

\subsection{Relativistic effects}
The electromagnetic field associated with a femtosecond laser pulse can be strong enough to induce relativistic effects, also contributing to the spin dynamics. Relativistic DFT and mean-field models based on the Dirac-Maxwell equations were developed in the past \cite{Keller, Engel, Romaniello}, but they are in general rather complex to handle either analytically or numerically.
More tractable models can be obtained by expanding the Dirac Hamiltonian in powers of $1/c$ \cite{Foldy, Hinsch,anant}. Second-order effects include the spin-orbit coupling and the Darwin correction, which are crucial for the proper understanding of magneto-optical processes in nano-objects \cite{Bigot_annphys}. They also lead to extra polarization and magnetization terms in the charge density and current \cite{anant}.
Recent attempts at incorporating relativistic effects include a fluid model derived from the Dirac equation \cite{asenjo2011}, as well as various semi-relativistic approaches, both fluid \cite{omar-jens} and kinetic \cite{asenjo2012,zamanian_POP10}.

\section{Conclusions}
Solid-state metallic objects display many features similar to those observed in high-temperature plasmas, the most obvious example being electron oscillations near the plasma frequency. A fair amount of modeling can be performed using the semiclassical approaches well known in the plasma physics community, ranging from kinetic equations of the Vlasov type to fluid models. However, particularly for nanometer scale objects, the electron density is so large that quantum effects cannot be neglected, both in the particle {\em statistics} (Fermi-Dirac, which can be incorporated into the semiclassical Vlasov approach) and in the {\em dynamics} (leading to quantum evolution equations such as the Hartree or DFT equations).
Such high densities also imply that the electron gas in a metal is not collisionless and therefore e-e correlations should be taken into account. This is a complex problem that is still being investigated \cite{suraud}.

In addition, electrons possess not only an electric charge, but also a spin, which interacts with magnetic fields, both external and self-consistent. Phase-space models can be adapted to accommodate the spin effects in a fully quantum fashion, although this may be more subtle for hydrodynamic models.
Finally, for strong enough laser excitations, the electrons can be so violently accelerated by the laser fields that relativistic effects come into play, also contributing to the spin dynamics.

\section*{Acknowledgements}
We thank the Agence Nationale de la Recherche, project Labex ``Nanostructures in Interaction with their Environment", for financial support.

\end{document}